\documentclass[10pt]{article}
\usepackage{amsmath}
\usepackage{graphicx}
\usepackage{amsfonts}
\usepackage{amssymb}
\usepackage{epsf}
\usepackage{latexsym}

\def\80{\hspace{0.8in}}

\newcommand{\be}{\begin{enumerate}}
\newcommand{\ee}{\end{enumerate}}
\newcommand{\bi}{\begin{itemize}}
\newcommand{\ei}{\end{itemize}}
\newcommand{\bd}{\begin{description}}
\newcommand{\ed}{\end{description}}

\def\beq{\begin{equation}}
\def\eeq{\end{equation}}
\def\bea{\begin{eqnarray}}
\def\eea{\end{eqnarray}}
\def\foo{\footnote}

\def\tilde{\widetilde}

\def\pa{\partial}
\def\d{\textrm{d}}

\def\cr{\mbox{\scriptsize{\bf $\mbox{ } \times \mbox{ }$}}}

\def\sA{\mbox{\scriptsize A}} 
\def\sB{\mbox{\scriptsize B}}

\def\sD{\mbox{\scriptsize D}}

\def\sF{\mbox{\scriptsize F}}

\def\sI{\mbox{\scriptsize I}}

\def\sM{\mbox{\scriptsize M}} 
 
\def\sO{\mbox{\scriptsize O}}

\def\sS{\mbox{\scriptsize S}}

\def\sW{\mbox{\scriptsize W}}

\def\eph(B){\mbox{\scriptsize emergent(LMB)}}

\def\tA{\mbox{\tiny A}}

\def\fA{\mbox{\sffamily A}}

\def\fE{\mbox{\sffamily E}}

\def\fH{\mbox{\sffamily H}}
\def\fI{\mbox{\sffamily I}}

\def\fL{\mbox{\sffamily L}}

\def\fO{\mbox{\sffamily O}}

\def\fQ{\mbox{\sffamily Q}}
\def\fR{\mbox{\sffamily R}}

\def\fT{\mbox{\sffamily T}}
\def\fU{\mbox{\sffamily U}}
\def\fV{\mbox{\sffamily V}}

\def\sfA{\mbox{\sffamily{\scriptsize A}}}
\def\sfB{\mbox{\sffamily{\scriptsize B}}}

\def\sfD{\mbox{\sffamily{\scriptsize D}}}

\def\sfM{\mbox{\sffamily{\scriptsize M}}}

\def\sfT{\mbox{\sffamily{\scriptsize T}}}
\def\sfU{\mbox{\sffamily{\scriptsize U}}}

\def\b{\underline{b}}
\def\p{\underline{p}}
\def\q{\underline{q}}
\def\a{\underline{a}}
\def\A{\underline{A}}
\def\B{\underline{B}}
\def\E{\underline{E}}

\begin{document}
\begin{titlepage}
\vspace{.7in}
\begin{center}

{\large{\bf NEW INTERPRETATION OF VARIATIONAL PRINCIPLES FOR GAUGE}} 

\vspace{0.1in}

{\large{\bf THEORIES. I. CYCLIC COORDINATE ALTERNATIVE TO ADM SPLIT.}} 

\mbox{ } 

\vspace{.4in}

\large{\bf Edward Anderson}$^1$ 

\vspace{.2in}

\large{\em Peterhouse, Cambridge CB2 1RD}\normalsize

\vspace{.2in}

\large{and} \normalsize

\vspace{.2in}

\large{\em DAMTP, Centre for Mathematical Sciences, Wilberforce Road, Cambridge CB3 OWA.} \normalsize

\end{center}

\begin{abstract}

I show how there is an ambiguity in how one treats auxiliary variables in gauge theories including 
general relativity cast as 3 + 1 geometrodynamics.  
Auxiliary variables may be treated pre-variationally as multiplier coordinates or as the velocities 
corresponding to cyclic coordinates.  
The latter treatment works through the physical meaninglessness of auxiliary variables' values applying 
also to the end points (or end spatial hypersurfaces) of the variation, so that these are free rather 
than fixed.    
[This is also known as variation with natural boundary conditions.]   
Further principles of dynamics workings such as Routhian reduction and the Dirac procedure 
are shown to have parallel counterparts for this new formalism.  
One advantage of the new scheme is that the corresponding actions are more manifestly relational.  
While the electric potential is usually regarded as a multiplier coordinate and 
Arnowitt, Deser and Misner have regarded the lapse and shift likewise,  
this paper's scheme considers new {\it flux}, {\it instant} and {\it grid} variables 
whose corresponding velocities are, respectively, the abovementioned previously used variables. 
This paper's way of thinking about gauge theory furthermore admits interesting generalizations, 
which shall be provided in a second paper.  

\end{abstract}

\mbox{ }

\mbox{ }


\vspace{4in} 

\noindent $^1$ ea212@cam.ac.uk

\end{titlepage}

\section{Introduction}

Let (M, $g_{\Gamma\Delta}(x^{\Xi}))$ be the spacetime topology and metric in the coordinates $x^{\Xi}$.  
Arnowitt, Deser and Misner (ADM) \cite{ADM} (\cite{MTW} also has a good exposition) split spacetime with 
respect to a family of spatial hypersurfaces, $\Sigma_{\lambda}$.  
Under this split, $x^{\Xi}$ splits into $(\lambda, x^{\xi})$, where $\lambda$ a coordinate label time 
for each  spatial hypersurface in the split and  $x^{\xi}(\lambda)$ a coordinatization of that spatial 
hypersurface.  
Then ADM take $g_{\Gamma\Delta}$ to split as follows:    
\beq
g_{\Gamma\Delta} = 
\left(
\stackrel{\mbox{$\beta_{\mu}\beta^{\mu} - \alpha^2$}}{\beta_{\gamma}}
\stackrel{\mbox{$\beta_{\delta}$}}{h_{\gamma\delta}}
\right) \mbox{ } .
\eeq
Here, $h_{\gamma\delta}(\lambda, x^{\mu})$ is the {\it induced metric}, i.e. the metric induced by 
the spacetime metric $g_{\Gamma\Delta}$ on the spacelike $\lambda$-hypersurface.  
$\alpha(\lambda, x^{\mu})$ is the {\it lapse}, which relates the passage of coordinate time between 
infinitesimally adjacent surfaces $\Sigma_{\lambda}$ and $\Sigma_{\lambda + \d \lambda}$ to the 
passage of proper time $\tau$:  
\beq
\d\tau = \alpha(\lambda, x^{\gamma})d\lambda \mbox{ }. 
\eeq
$\beta(\lambda, x^{\mu})$ is the {\it shift}, which relates the coordinates on $\Sigma_{\lambda + 
\d \lambda}$ to those on $\Sigma_{\lambda}$: 
\beq
x^{\gamma}(\lambda + \d \lambda) - x^{\gamma}(\lambda) = - \beta^{\gamma}\d \lambda \mbox{ } .  
\eeq
I.e., the lapse is the change in spacing in how the spatial hypersurfaces are stacked, while the shift 
is the displacement made in identifying the spatial coordinates of one slice with those on an adjacent 
slice.
Thus these two can be envisaged together as the determiners of what extrinsic geometry each spatial 
hypersurface $\Sigma_{\lambda}$ is to have within spacetime (see e.g. p 346 of \cite{WheelerGRT}).     
The ADM split is then applied to the spacetime form of the general relativity (GR) action to produce an 
ADM action for split space-time,\foo{
I use
( ) for function dependence, 
[ ] for functional dependence,
( ; ] for a mixture of function dependence before the semi-colon and functional dependence after it.
I use lower-case Greek letters for spatial indices, except that $\alpha$ and $\beta$ are reserved 
to denote lapse and shift.   
$h = h(h_{\mu\nu}(x^{\pi}))$, $D_{\mu}$, and $R(x^{\pi}; h_{\mu\nu}(x^{\pi})]$ are, 
respectively, the determinant, covariant derivative and Ricci scalar associated with the 3-metric 
$h_{\mu\nu}(x^{\pi})$.  
The dot denotes ${\pa}/{\pa\lambda}$ from Sec 4 onwards.  
$\pounds_{\beta}$ is the Lie derivative with respect to $\beta_{\gamma}$.  
$G_{\alpha\beta\gamma\delta} = 
\frac{1}{\sqrt{h}}
\left\{
 h_{\mu\nu}h_{\rho\sigma} - \frac{1}{2}h_{\mu\nu}h_{\rho\sigma}
\right\}$ 
is the DeWitt supermetric \cite{DeWitt} with inverse 
${G}^{\mu\nu\rho\sigma} = \sqrt{h}
\{  h^{\mu\rho}h^{\nu\sigma}  -  h^{\mu\nu}h^{\rho\sigma}   \}$ which plays the role of kinetic metric 
(see e.g. \cite{Lanczos}) in GR.  
Overbar denotes densitization (multiplication by $\sqrt{h}$) and underbar denotes dedensitization 
(division by $\sqrt{h}$).} 
\beq
\fI_{\sA\sD\sM}[h_{\mu\nu}, \dot{h}_{\mu\nu}, \beta_{\mu}, \alpha] = 
\int\d\lambda\int\d^3x\bar{\fL}
(\dot{h}_{\mu\nu}, \alpha; h_{\mu\nu}, \beta_{\mu}]  
=  \int\d\lambda\int\d^3x\sqrt{h}\alpha 
\left\{
\frac{\underline{G}^{\mu\nu\rho\sigma}\{\dot{h}_{\mu\nu} - \pounds_{\beta}h_{\mu\nu}\}
\{\dot{h}_{\rho\sigma} - \pounds_{\beta}h_{\rho\sigma}\}}{4{\alpha}^2} + R 
\right\} \mbox{ } .
\eeq
The ADM split and variants have important applications in quantum gravity (see e.g. \cite{DeWitt, 
Battelle, I93, Kuchar}) and in the study of the compact binary problem in numerical general relativity 
(see e.g. \cite{BSG}).

It is `standard lore' in varying the ADM action to regard the lapse and the shift as coordinates;  
as the corresponding velocities do not feature in the action, these are Lagrange multiplier coordinates.  
Denoting Lagrange multiplier coordinates in general by $m_{\sfB}$, variation with respect to these   
produces simplified Euler--Lagrange equations of the form 
\beq
\frac{\pa\fL}{\pa m_{\sfB}} = 0 \mbox{ } .  
\label{timothi}
\eeq
This is how the Hamiltonian and momentum constraints of GR arise in the ADM formulation (see the 
Appendix).

Now, it is a fairly standard observation (see e.g. \cite{Thorne}) that the above expressions for the 
lapse and shift can, at the post-variational level, be straightforwardly rearranged to be interpreted 
as velocities, 
\beq
\alpha(\lambda, x^{\gamma}) = \frac{\d \tau}{\d \lambda} \mbox{ } , \mbox{ } 
\beta^{\gamma}(\lambda, x^{\gamma}) = - \frac{\d x^{\gamma}}{\d \lambda} \mbox{ } .  
\eeq   
What's new in this paper is to regard the objects of which the lapse and the shift are the velocities 
as new variables {\sl prior} to variation.  
Reading off (\ref{BFOA}), these are the proper time $\tau$ attributed to the spatial slice in question 
and the grid of coordinates $x^{\mu}$ on that slice, but in their capacity as new variables to 
be varied I term these the {\it instant} $I(\lambda, x^{\mu})$ and the {\it grid} 
$F^{\Gamma}(\lambda, x^{\mu})$. 
[I use an `$F$' to denote this as it is a subcase of a more widely applicable concept of spatial 
{\it frame} variable].    
Through showing this to work out acceptably as an alternative interpretation, this paper shows that 
there is in fact an ambiguity in the long-standing, well-known and much-used statement that the lapse 
and shift are Lagrange multipliers

Moreover, the alternative better complies with relational first principles, which is of interest 
along the lines of establishing GR to have multiple foundations, along the lines of Wheeler's arguments 
for `many routes to GR' \cite{Battelle, MTW, HKT}.\footnote{For, 
different routes offer different techniques and different insights, suggest different alternatives 
or generalizations against which the established theory can be tested, and different routes to a given 
classical theory may be quantum-mechanically inequivalent.}  
%
The relational perspective, implements ideas of Leibniz \cite{Leibniz} and Mach \cite{Mach} to modern 
physics along the lines of Barbour's work \cite{BB82, B94I, RWR}.    
One starts with a configuration space $\fQ$ of (models of) whole-universe systems.

Then {\it configurational relationalism} is the declaration that certain motions acting on $Q$ are to 
be physically meaningless.  
One way\footnote{This is along the lines of \cite{Lan}, which   
is closely connected with well-known approaches to gauge theory. 
For Barbour's own and somewhat conceptually different way of thinking about configurational 
relationalism (`best matching'), see e.g. \cite{BB82, B03}.}
of implementing this  is to use arbitrary-$G$-frame-corrected quantities rather than 
bare $\fQ$ configurations, where $G$ is the group of physically meaningless motions.  
Despite this augmenting $\fQ$ to $\fQ \times G$, variation with respect to each adjoined independent 
auxiliary $G$-variable produces a constraint which wipes out one $G$ variable and one redundancy among 
the $\fQ$ variables, so that one ends up on the quotient space $\fQ/G$ (the desired reduced 
configuration space).  
This is widely a necessity in theoretical physics through working on the various reduced spaces directly 
often being technically unmanageable.

{\it Temporal relationalism} is the notion that there is no meaningful primary notion of time for 
the universe as a whole.  
One implementation of temporal relationalism in modern physics is through using manifestly 
reparametrization invariant actions, though Barbour would prefer this to be attained without 
any extraneous time variables either.   
If one is working towards deriving GR in this picture, one would e.g. take $\fQ$ to be the space of 
positive-definite 3-metrics on a fixed spatial topology, Riem($\Sigma$), and $G$ to be the 
3-diffeomorphisms Diff($\Sigma$) that correspond to the coordinatization of space being physically 
meaningless.   
It is then clear upon inspecting the split space-time GR action that the instant and grid 
interpretations of the auxiliaries therein give the action a manifest reparametrization invariance 
while the lapse and shift multiplier coordinate interpretations obstruct this by inhomogenizing 
the Lagrangian's dependence on $\pa/\pa\lambda$.
To have no extraneous time variables either, one would require not an ADM-type action with its 
manifest lapse or velocity of the instant, but a Baierlein--Sharp--Wheeler (BSW) \cite{BSW} 
or BFO-A\footnote{This 
action (\ref{BFOA}) first appeared in \cite{Lan}, building on \cite{RWR} and insights in \cite{ABFO}.  
See also \cite{Phan} for discussion.}  
type action in which this has been eliminated 
(or, alternatively, such an action could be adopted as one's starting-point).  
This paper completes this work by mapping out how the instant can be eliminated from the instant--grid 
version of the split GR action so as to produce the BFO-A action.

There is an obstruction that explains why this kind of alternative interpretation of auxiliary 
variables has not been spotted until recently.
At first sight, it looks preposterous to treat the lapse and shift as velocities prior to variation.  
This is because cyclic coodinates $c_{\sfB}$ are well-known to lead to simplified Euler--Lagrange 
equations of the form\foo{In 
Secs 1--3 the dot denotes $\d/\d\lambda$.} 
\beq 
\frac{\delta\fL}{\delta \dot{c}_{\sfB}} = C^{\sfB} \mbox{ } , \mbox{ constant} \mbox{ } ,
\label{cycliceq}
\eeq
which are generally not equivalent to the (\ref{timothi}) produced by the preceding formalism.  
However, there is a subtlety. 
In a gauge formulation of a physical theory, the configuration variables contain both dynamical degrees 
of freedom and nondynamical/auxiliary variables.  
Then by the gauge principle (or configurational relationalism), the reference frame variables are to be 
taken to be unphysical, i.e. arbitrarily specifiable at each instant, with no measurable quantities 
depending on which choice of these is made.  
This includes among the things that are to be arbitrary the values of the auxiliary arbitrary G-frame 
variables at the end points (or end spatial hypersurfaces in the case of field theory) of the variation.  
Thus it is free end point variation (also known as variation with natural boundary conditions) 
\cite{CHFoxBM} that turns out to be the correct type of variation as regards reference frame 
variables in gauge theories.  
In Sec 2, I show that this in no way affects the outcome of varying with respect to multiplier 
variables, but it does alter the outcome of varying with respect to cyclic variables precisely so as to 
remove the above inequivalence.

In Sec 3, I then show that there is also no difference between multiplier elimination in the auxiliary 
multiplier formulation and cyclic velocity elimination (by the more subtle procedure of Routhian 
reduction, see e.g. \cite{Lanczos}) in the above auxilary cyclic coordinate formulation.  
I also explain how one subsequently gets an alternative to Dirac's appending procedure \cite{Dirac}.     
These sections are supported by relational particle mechanics toy models \cite{BB82, B94I, EOT, ERPM, 
ERPMSRPM, SRPM} -- zero total angular momentum universes formulated by frame-correcting with respect to 
the rotations with a rotational auxiliary analogue of the grid--shift object.  
Armed with all this, in Sec 4 I give the manifestly reparametrization-invariant alternative to the ADM 
formulation of Einstein--Maxwell theory.  
This requires, in addition to the lapse and shift being re-interpreted pre-variationally as instant and 
grid variables, for the conventional electric potential $\Phi$ to be re-interpretated as the velocity 
of a new (corrected magnetic) {\it flux} variable $\Psi$ (which is another example of frame variable).    
Then performing Routhian reduction on this action so as to eliminate the instant leads to the recovery 
of the BFO-A action that complies with the first principles of the relational approach, which working is 
in direct parallel to how multiplier elimination of the lapse from the ADM action gives the BSW action.    
I conclude in Section 5 with a discussion of applications and extensions, among which the case of 
auxiliaries more general than multipliers or cyclic coordinates is presented in a subsequent paper.  
The Appendix provides Einstein--Maxwell theory's ADM and BSW formulations (and the BSW elimination 
that links them) for useful comparison.

\section{An alternative picture for gauge formulations}

One common way of formulating gauge theories is in the Lagrangian picture, in which one builds 
Lagrangians that possess certain groups of symmetries, $G$.  
Another common way of formulating gauge theories is in the Hamiltonian picture, in which one appends 
certain constraints (e.g. in the sense of Dirac \cite{Dirac}) that are related to the action of the 
gauge group.    
The Dirac procedure has the additional advantage that even if one does not know what symmetries 
the theory is to possess, the procedure nevertheless systematically derives the theory's constraints 
(see e.g. \cite{HTbook}).  
One can move between the two pictures by Legendre transformation.  
One approach within the Lagrangian picture is to attempt to {\sl impose} a group $G$ of symmetries on a 
Lagrangian, and then use the Dirac procedure to systematically determine whether this imposition is 
consistent and complete.\foo{Constraints
that do not correspond to $G$ could arise as integrabilities for other constraints in the 
theory, signalling that $G$ need be extended if one is to get a consistent theory.  
See the end of Sec 4 for an example of this.  
So many constraints can arise in this way that the resulting ``theory" ends up trivial (all degrees of 
freedom used up) or inconsistent (more independent constraints than degrees of freedom) \cite{RWR}.}

One way of building a formulation in which the group of symmetries $G$ (whether actual, expected or 
imposed) is present and manifest is to enlarge the configuration space $\fQ$ to $\fQ \times G$ by 
adjoining auxiliary variables $g_{\sfB}$.    
These show up in the Lagrangian formalism as arbitrary G-frame corrections to the $\fQ$-variables 
\cite{Lan, Phan}.    
To see how this gives gauge theories, consider for the moment that the auxiliary variables take their 
usual guise as multipliers in the generalized sense (coordinates whose velocities do not occur in the 
action). 
E.g. one conventionally views thus the electric potential $\Phi$, and 
the shift $\beta^{\gamma}$ and lapse $\alpha$ of GR split with respect to spatial hypersurfaces.     
Further examples are the translation and rotation auxiliaries, $\a$ and $\b$ in mechanics models that 
are spatially relational with respect to the Euclidean group \cite{BB82, ERPM, ERPMSRPM} (as opposed to 
being based on the supposition of absolute space).

Variation with repect to the auxiliaries $g_{\sfB}$ then produces 
\beq 
\frac{\delta\fL}{\delta g_{\sfB}} = 0 \mbox{ } ,  
\label{multipliereq}
\eeq
which are the constraints that one is expecting if one is indeed to obtain thus a gauge theory with 
gauge group $G$.  
These constraints then use up both the auxiliary variables' degrees of freedom and an equal amount
of degrees of freedom from the $\fQ$, thus indeed leaving one with a theory in which the dynamical 
variables pertain to the quotient space $\fQ/G$ (i.e. variables among the original $\fQ$ which are 
entirely unaffected by which choice of $G$-frame is made).

As a simple example, consider the translation and rotation invariant mechanics \cite{BB82, ERPM, 
ERPMSRPM}  that follows from the Jacobi-type action 
\beq
\fI[\q_I, \dot{\q}_I, {\a}, {\b}] = \int\d\lambda \fL(\q_I, \dot{\q}_I, {\a}, {\b}) = 
2\int\d\lambda\sqrt{\fT\{\fV + \fE\}}
\eeq
with kinetic term
\beq 
\fT(\q_I, \dot{\q}_I, {\a}, {\b}) = \frac{1}{2}\sum_{I = 1}^N m_I
|\dot{q}_I - {\a} - {\b} \cr \q_I|^2 \mbox{ } ,
\eeq
potential term 
\beq 
\fV = \fV(|r_{IJ}| \mbox{alone}) \mbox{ } 
\eeq   
and total energy $\fE$.
Then variation with respect to $\a$ gives the multiplier equation 
$0 = {\pa\fL}/{\pa \a} = \sum_{I = 1}^{N}\p_I$, i.e. zero total momentum, 
and variation with respect to $\b$ gives the multiplier equation 
$0 = {\pa\fL}/{\pa \b} = \sum_{I = 1}^{N}\q_I \cr \p_I$, i.e. zero total angular momentum.
See the Appendix for another example (the standard ADM split of Einstein--Maxwell theory).

The main point of the present Paper, however, is that taking $\a$, $\b$, $\Phi$, $\beta^{\gamma}$ and 
$\alpha$ to be multipliers is not the only viable interpretation.  
They may also be interpreted as velocities associated with an auxiliary cyclic coordinate.  
This is possible because the variables in question are {\sl auxiliary}.  
Thus by the gauge principle, their values are entirely arbitrary.  
Thus, in particular, they take arbitrary rather than fixed values at the end points of the variation.    
Thus the appropriate type of variation is {\it free end point (FEP) variation} 
(also known as {\it variation with natural boundary conditions}) \cite{CHFoxBM}.
FEP variation involves more freedom than standard variation, i.e. it involves a larger space 
of varied curves,   
\beq
q_{\sfA}(\lambda, \mu, \nu) = \tilde{q}_{\sfA}(\lambda) + \mu Y(\lambda) + \nu Z(\lambda) 
\mbox{ } \mbox{ such that } \mbox{ } 
Y(\lambda_i) = 0 \mbox{ } \mbox{ and } \mbox{ } Z(\lambda_f) = 0 \mbox{ }   
\eeq 
(where $\mu$, $\nu$ are further parameters).  
Then consider configurations $\{q_{\sfA}, g_{\sfB}\}$ for $g_{\sfB}$ entirely arbitrary through being 
unphysical.  
With this in mind, I start from scratch, with a more general Lagrangian in which both the auxiliaries 
$g_{\sfB}$ and their velocities $\dot{g}_{\sfB}$ in general occur: 
$\fL(\lambda, q_{\sfA}, g_{\sfB}, \dot{q}_{\sfA}, \dot{g}_{\sfB})$.

Then variation with respect to $g_{\sfB}$ gives  
\beq 
0 = \delta \fI = \int_{\lambda_i}^{\lambda_f}\d\lambda\delta 
\fL(\lambda, q_{\sfA}, g_{\sfB}, \dot{q}_{\sfA}, \dot{g}_{\sfB}) =
\int_{\lambda_i}^{\lambda_f}\d\lambda\delta g_{\sfB} 
\left\{ 
\frac{\pa\fL}{\pa g_{\sfB}} - \frac{\d}{\d\lambda} 
\left\{ 
\frac{\pa \fL}{\pa \dot{g}_{\sfB}}
\right\}
\right\} + 
\left\{ 
\frac{\pa\fL}{\pa \dot{g}_{\sfB}}\delta g_{\sfB}
\right\}_{\lambda_i}^{\lambda_f} \mbox{ } .  
\label{once}
\eeq
Because $g_{\sfB}$ is auxiliary, this variation is free, so $\delta g_{\sfB}|_{\lambda_i}$ and 
$\delta g_{\sfB}|_{\lambda_f}$ are not controllable. 
Thus one obtains 3 conditions per variation, 
\beq 
\frac{\pa\fL}{\pa g_{\sfB}} = \frac{\d}{\d\lambda} 
\left\{ 
\frac{\pa\fL}{\pa \dot{g}_{\sfB}} \right\} 
\mbox{ , }
\left.
\frac{\pa\fL}{\pa \dot{g}_{\sfB}}
\right|_{\lambda_i} = 0 = 
\left.
\frac{\pa\fL}{\pa \dot{g}_{\sfB}}
\right|_{\lambda_f} \mbox{ } , 
\label{doce}
\eeq 
or, in terms of momenta $p^{\sfB} \equiv {\pa \fL}/{\pa \dot{g}_{\sfB}}$, 
\beq
\frac{\pa\fL}{\pa g_{\sfB}} = \dot{p}^{\sfB} \mbox{ , }
\left.
p^{\sfB}
\right|_{\lambda_i} 
= 0 = 
\left. 
p^{\sfB}
\right|_{\lambda_f} 
\mbox{ } .
\eeq

If the auxiliaries $g_{\sfB}$ are multipliers $m_{\sfB}$, (\ref{once}) reduces to
\beq 
p_{\sfB} = 0 \mbox{ } , \mbox{ } \frac{\pa\fL}{\pa m_{\sfB}} = 0
\eeq
and redundant equations.
I.e. the $\lambda_i$ and $\lambda_f$ terms automatically vanish in this case by applying the multiplier 
equation to the first factor of each.  
This is the case regardless of whether the multiplier is not auxiliary and thus standardly varied, or
auxiliary and thus FEP varied, as this difference translates to whether or not the cofactors 
of the above zero factors are themselves zero or not.  
Thus the FEP subtlety in no way affects the outcome in the multiplier coordinate case.

If the auxiliaries $g_{\sfB}$ are cyclic coordinates $c_{\sfB}$, the above reduces to 
\beq 
\left.
p^{\sfB}
\right|_{\lambda_i} 
= 0 = 
\left.
p^{\sfB}
\right|_{\lambda_f} \mbox{ } 
\label{ckill}
\eeq
and
\beq 
\dot{p}^{\sfB} = 0 \mbox{ }  , 
\label{hex}
\eeq
which implies that 
\beq 
p^{\sfB} = C \mbox{ } ,
\eeq 
but $C$ is identified as 0 at either of the two end points (\ref{ckill}), and, being constant, 
is therefore zero everywhere.  
Thus (\ref{hex}) and the definition of momentum give 
\beq 
\frac{\pa \fL}{\pa \dot{c}_{\sfB}} \equiv p^{\sfB} = 0  \mbox{ } .  
\eeq
So, after all, one does get equations that are equivalent to the multiplier equations 
(\ref{multipliereq}) at the classical level, albeit there are conceptual and foundational reasons to 
favour the latter, as laid out in the Introduction.  
For previous discussion of this case in the literature, see \cite{SRPM, ABFO, Lan, ABFKO}.

For the translation and rotation invariant mechanics example, this works out as follows.  
The appropriate action is 
\beq
\fI[\q_I, \dot{\q}_I, \dot{\a}, \dot{\b}] = \int\d\lambda \fL(\q_I, \dot{\q}_I, \dot{\a}, \dot{\b}) = 
2\int\d\lambda\sqrt{\fT\{\fU + \fE\}} \mbox{ } , 
\eeq
with
\beq 
\fT(\q_I, \dot{\q}_I, \dot{\a}, \dot{\b}) = \frac{1}{2}\sum_{I = 1}^Nm_I
|\dot{q}_I - \dot{\a} - \dot{\b} \cr \q_I|^2 
\eeq
and $\fU = - \fV$.  
Then variation with respect to $\a$ gives $C = {\pa\fL}/{\pa \dot{\a}} = \sum_{I = 1}^{N}\p_I$ and the 
FEP conditions $\pa\fL/\pa \dot{\a}|_{\lambda_i} = 0 = \pa\fL/\pa \dot{\a}|_{\lambda_f}$, 
so $C = \pa\fL/\pa \dot{\a} = 0$ at the end point, but $C$ is constant, so $C = 0$ everywhere, so  
one obtains $\sum_{I = 1}^{N}\p_I = 0$ again. 
And variation with respect to $\b$ gives $D = {\pa\fL}/{\pa \dot{\b}} = \sum_{I = 1}^{N}\q_I \cr \p_I$ 
and FEP conditions $\pa\fL/\pa \dot{\b}|_{\lambda_i} = 0 = \pa\fL/\pa \dot{\b}|_{\lambda_f}$, 
so $D = \pa\fL/\pa \dot{\b} = 0$ at the end point, but $D$ is constant, so $D = 0$ everywhere, so  
one obtains $\sum_{I = 1}^{N}\q_I \cr \p_I = 0$ again. 
For further examples of this kind of approach to gauge theories, see e.g. Sec. 4 (3 + 1 split GR), 
Sec. 5 (3 + 1 split of Einstein--Maxwell theory),  \cite{SRPM, ERPMSRPM} (mechanics that is 
scale-invariant as well as translation and rotation invariant), and \cite{CG, ABFO, ABFKO} alongside 
Paper II for a conformal theory of gravity and conformogeometrodynamical formulations of GR 
(conformal split of 3 + 1 split GR along the lines of York's work \cite{York} on the 
initial-value formulation).

The case of (\ref{doce}) in which the auxiliary is neither cyclic nor a multiplier I leave for Paper II 
\cite{ADMII}, as the conformogeometrodynamical formulations in which it applies \cite{ABFO, ABFKO}  
are rather more complicated than the present paper's examples.

As regards how one determines in the first place which variables are the auxiliaries that one is to 
FEP vary with respect to, it turns out that for the type of examples in this suffices 
to FEP vary with respect to the $g_{\sfB}$, as follows.    
While some sets of variables contain partly physically relevant and partly physically 
irrelevant degrees of freedom, the gauge principle only gives license to vary in the FEP way 
with respect to sets of variables that are purely physically irrelevant.  
So FEP variation should be applied only to isolated physically irrelevant variables.  
Moreover, one does not have to isolate all of these, as follows.  
In gauge theory, physically irrelevant variables come in pairs that are associated with a single 
constraint, which, when taken onto account, ensures that neither member of the pair remains in the 
theory's equations.    
Variation with respect to either member of the pair produces the same constraint.
Thus to get all of the constraints, one needs to isolate a full half-set of auxiliary variables i.e. 
to isolate one representative from each of the above pairs.  
But in the arbitrary $G$-frame method as used in this paper, the frame variables $g_{\sfB}$ that one 
adjoins are clearly both such a half-set and already-separate. 
Thus for the type of theories considered in this paper, it suffices to FEP vary with respect 
to the $g_{\sfB}$.

\section{Establishing further equivalences between auxiliary cyclic coordinates and auxiliary multiplier 
coordinates}

We also need to establish that the a priori distinct procedures of cyclic velocity elimination (known as 
Routhian reduction) and muliplier elimination are equivalent.  
Consider $\fL(q_{\sfA}, \dot{q}_{\sfA}, g_{\sfB})$ where $g_{\sfB}$ are auxiliary coordinates.  
If these are taken to be multiplier coordinates $m_{\sfB}$, then variation yields 
$0 = {\pa\fL}/{\pa m_{\sfB}}$.  
If this is soluble for $m_{\sfB}$, one can replace it by $m_{\sfB} = m_{\sfB}(q_{\sfA}, \dot{q}_{\sfA})$, 
and then substitute that into $\fL$.  
If the $g_{\sfB}$ are taken to be velocities corresponding to cyclic coordinates $c_{\sfB}$, 
then FEP variation yields 
\beq
0 = \frac{\pa\fL}{\pa \dot{c}_{\sfB}} = p_{\sfB} = {\cal C}_{\sfB} \mbox{ } . 
\label{sku}
\eeq
This is soluble for $\dot{c}_{\sfB}$ iff the above is soluble for $m_{\sfB}$.  
However one now requires passage to the Routhian in eliminating $\dot{c}_{\sfB}$ from the action: 
$\fR(q_{\sfA}, \dot{q}_{\sfA}) = \fL(q_{\sfA}, \dot{q}_{\sfA}, \dot{c}_{\sfA}) - \dot{c}_{\sfB}p^{\sfB}$.  
But the last term is zero, by (\ref{sku}), so Routhian reduction in the cyclic velocity interpretation 
is equivalent to multiplier elimination in the multiplier coordinate interpretation.  


Next, I consider how the Dirac procedure implies encodement by auxiliaries at the level of the 
Lagrangian.  
From the bare (G-uncorrected Lagrangian) $\fL^{\mbox{\scriptsize bare}}(q_{\sfA}, \dot{q}_{\sfA})$, 
Legendre transformation gives  $\fH(q_{\sfA}, p^{\sfA}) = \dot{q}_{\sfA}p^{\sfA} - 
\fL^{\mbox{\scriptsize bare}}(q_{\sfA}, \dot{q}^{\sfA})$ for which ${\cal C}_{\sfB} = 0$ are discovered 
by the first part of the Dirac procedure.  
Then, in the multiplier interpretation,  
\beq 
\fH(q_{\sfA}, p^{\sfA}, m_{\sfB}) = \dot{q}_{\sfA}p^{\sfA} - 
\fL^{\mbox{\scriptsize bare}}(q_{\sfA}, \dot{q}^{\sfA}) + \mbox{m}_{\sfB}{\cal C}^{\sfB}
\eeq
so that this procedure standardly follows from the standard Dirac procedure of appending constraints 
multiplied by Lagrange multipliers.  
In the cyclic velocity interpretation however, one has a new alternative to Dirac appending that involves appending constraints multiplied by 
velocities of cyclic coordinates.  
Then one is not forming a Hamiltonian, but rather an `almost-Hamiltonian', in that it depends not only 
on positions and momenta but also on the velocities of the auxiliary cyclic coordinates, 
\beq
\fA(q_{\sfA}, p^{\sfA}, \dot{c}^{\sfB}) = \dot{q}_{\sfA}p^{\sfA} + \dot{c}_{\sfB}{\cal C}^{\sfB} - 
\fL^{\mbox{\scriptsize bare}}(q_{\sfA}, \dot{q}_{\sfA}) \mbox{ } .  
\eeq


There is no problem uplifting Sec 2 or 3 to field theory.  
[Now one has free end {\sl spatial hypersurfaces} (FESH) and $C(x^{\mu})$ in place of $C$, constant, but 
all the above arguments straightforwardly carry through].

\section{Manifestly reparametrization invariant alternatives to the ADM and BSW-type formulations of 
Einstein--Maxwell theory}

As delineated in the Introduction and Appendix, in the ADM split, lapse and shift variables are used and 
have the status of Lagrange multipliers.
In this paper, instead, I use the new split
\beq
g_{\Gamma\Delta} = 
\left(
\stackrel{\mbox{$\dot{F}_{\rho}\dot{F}^{\rho} - \dot{I}^2$}}{\dot{F}_{\mu}} \mbox{ }
\stackrel{\mbox{$\dot{F}_{\nu}$}}{h_{\mu\nu}}
\right)
\eeq
where $F_{\mu}$ is the frame (grid) and $I$ is the instant, the meanings of which are explained in the 
Introduction.    
The considerations in Sec 2 shall ensure that the subsequent variation works out.  

The GR kinetic term is now 
\beq
\fT = \underline{G}^{\mu\nu\rho\sigma}\{\dot{h}_{\mu\nu}   -   \pounds_{\dot{F}}h_{\mu\nu}\}
                                      \{\dot{h}_{\rho\sigma} - \pounds_{\dot{F}}h_{\rho\sigma}\}
\mbox{ } .  
\eeq 
and the GR potential term is $R$ as before.

For greater generality, I include the electromagnetic field (massless spin 1 field) as an example 
of matter field.\footnote{
Including scalar fields is straightforward, and inclusion of spin-1/2 fermions 
follows along standard lines.  This paper's treatment of electomagnetism generalizes to Yang--Mills 
theory and to the various associated scalar and spin-1/2 gauge theories.}
%
In Maxwell electrodynamics, the usually-employed electric potential variable is defined by  
\beq
\underline{\nabla}\Phi = - \underline{E} - \underline{\dot{A}} \mbox{ } .  
\label{std}
\eeq
Here $\A$ is the magnetic potential such that $\B = \underline{\nabla} \cr \A$, where $\B$ is the 
magnetic field and $\E$ is the electric field.  
The variable I use in this paper is, rather, $\Psi$ such that $\dot{\Psi} = \Phi$.  
In the same way that $\Phi$ is a `work per unit charge' $\int_{\Gamma(t)}\frac{F_{\mu} \d x^{\mu}}{q}$, 
$\Psi$ is an `action per unit charge' $\int \d t \int_{\Gamma(t)}\frac{F_{\mu} \d x^{\mu}}{q}$.   
Another interpretation is as follows.  
In the flat spacetime context, integrating (\ref{std}) and applying the Faraday--Lenz law to the first 
factor and Stokes's theorem, mixed partial equality and the definition of $\underline{A}$ to the second 
factor, 
\beq
\Phi = 
- \oint_{\Gamma(t)}\{E_{\mu} + A_{\mu}\}dx^{\mu} = 
\frac{\d}{\d t}\int\int_{S(t)}\underline{B}\cdot\underline{\d S} -  
\int\int_{S(t)}\underline{\dot{B}}\cdot\underline{\d S} 
= \left( 
\stackrel{\mbox{a corrected rate of change}}{\mbox{of magnetic flux}}
\right)
\mbox{ }   
\eeq
(for surface $S$ with boundary curve $\Gamma$).  
By Leibniz's rule and theorem this is also 
\beq
\Phi = \int\int_{\frac{\pa S(t)}{\pa t}}\underline{B}\cdot\underline{\d S} = 
\left( 
\stackrel{\mbox{change in magnetic flux due to}}{\mbox{change in shape of the surface over time}}
\right) 
\mbox{ } .  
\eeq
Thus  
\beq
\Psi = \int\int_{S(t)}\underline{B}\cdot\underline{\d S} -  
\int\d t \int\int_{S(t)}\underline{\dot{B}}\cdot\underline{\d S} = 
\left( 
\mbox{corrected magnetic flux}
\right) \mbox{ } , 
\eeq
the first term being magnetic flux and the second term being the correction; this is also  
\beq
\Psi = \int\d t\int\int_{\frac{\pa S(t)}{\pa t}}\underline{B}\cdot\underline{\d S} = 
\left( 
\stackrel{\mbox{magnetic flux due to change in shape}}
         {\mbox{of the surface over time integrated over time}}       
\right) 
\mbox{ } .  
\eeq

Electromagnetism in this formulation has kinetic and potential terms  
\beq 
\fT(\dot{A}_{\mu}; \dot{\Psi}] = \frac{1}{2}| \dot{A}_{\mu} - \pa_{\mu}\dot{\Psi}|^2 
\mbox{ } , \mbox{ } 
\fV[A_{\mu}] = \frac{1}{4}F_{\mu\nu}F^{\mu\nu}
\eeq 
for $F_{\mu\nu} = \pa_{\mu}A_{\nu} - \pa_{\nu}A_{\mu}$ the spatial part of the electromagnetic 
field strength tensor.  
This formulation does indeed work in the flat spacetime context: $C(x^{\mu}) = 
{\delta\fL}/{\delta\dot{\Psi}} = \pa_{\mu}\Pi^{\mu}$ but $C$ on the end spatial hypersurface is 0 and 
$C$ is time independent, so $C$ is 0 everywhere, and thus we recover the Gauss constraint, 
$\pa_{\mu}\Pi^{\mu} = 0$.

I then obtain the suitable manifestly reparametrization invariant action for Einstein--Maxwell theory 
by building the above up to hold on curved spaces and then coupling it to GR so that it holds on 
dynamically changing curved spaces.  
The complete Einstein--Maxwell action is then  
$$
\fI_{\sfA}[h_{\mu\nu}, \dot{h}_{\mu\nu}, A_{\mu}, \dot{A}_{\mu}, \dot{F}_{\mu}, \dot{I}, \dot{\Psi}] = 
\int\d\lambda\int\d^3x\dot{I}\bar{\fL}_{\sfA} 
(\dot{h}_{\mu\nu}, \dot{A}_{\mu}, \dot{I}; h_{\mu\nu}, A_{\mu}, \dot{F}_{\mu}, \dot{\Psi}] \equiv  
\int\d\lambda\int\d^3x\sqrt{h}\dot{I}
\left\{
\frac{\fT}{4\dot{I}^2} + \frac{\fT_{\sA}}{\dot{I}^2} + R - \fV_{\sA}
\right\}
$$
$$
=  \int\d\lambda\int\d^3x\sqrt{h}\dot{I}
\left\{
\frac{\underline{G}^{\mu\nu\rho\sigma} 
\{\dot{h}_{\mu\nu} - \pounds_{\dot{F}}h_{\mu\nu}\}
\{\dot{h}_{\rho\sigma} - \pounds_{\dot{F}}h_{\rho\sigma}\}}{4\dot{I}^2} + 
\frac{        h^{\mu\nu}\{\dot{A}_{\mu} - \pounds_{\dot{F}}A_{\mu} - \pa_{\mu} \dot{\Psi}\}
                        \{\dot{A}_{\nu} - \pounds_{\dot{F}}A_{\nu} - \pa_{\nu} \dot{\Psi}\}        }
     {       2\dot{I}^2        } \right.
$$
\beq
\left.  
- \frac{1}{4}F_{\mu\nu}F^{\mu\nu} + R \right\}
\label{ADMsplit} 
\eeq

From this starting point, the conjugate momenta are 
\beq
\pi^{\mu\nu} = \frac{\delta \bar{\fL}_{\sA}}{\pa \dot{h}_{\mu\nu}} = \frac{\sqrt{h}}{2\dot{I}}
G^{\mu\nu\rho\sigma}
\{\dot{h}_{\rho\sigma} - \pounds_{\dot{F}}h_{\rho\sigma}\} \mbox{ }  
\mbox{ } \mbox{ and } \mbox{ }
\Pi^{\mu} \equiv 
\frac{\pa {\fL}}{\pa \dot{A}_{\mu}} = 
\frac{\sqrt{h}}{\dot{I}}\{\dot{A}^{\mu} - \pounds_{\beta}A^{\mu} - \pa^{\mu}\dot{\Psi}\} 
\mbox{ } . 
\eeq
Variation with respect to $I$ gives as a secondary constraint
\beq
C(x^{\mu}) = \frac{\pa\bar{\fL}_{\sA}}{\pa\dot{I}} = G_{\mu\nu\rho\sigma}\pi^{\mu\nu} \pi^{\rho\sigma} + \frac{1}{2\sqrt{h}}\Pi_{\mu}\Pi^{\mu} + 
\frac{\sqrt{h}}{4} F_{\mu\nu} F^{\mu\nu}  - \sqrt{h}R   
\mbox{ } ,   
\label{Ham}
\eeq
and the FESH conditions $\pa\bar{\fL}_{\sA}/\pa\dot{I}|_{\lambda_i} = 0 = 
\pa\bar{\fL}_{\sA}/\pa\dot{I}|_{\lambda_f}$, 
so $C(x^{\mu}) = \pa\bar{\fL}_{\sA}/\pa\dot{I} = 0$ at the end spatial hypersurface, but 
$C$ is hypersurface-independent, so $C = 0$ everywhere.  
Thus one arrives at 
\beq
G_{\mu\nu\rho\sigma}\pi^{\mu\nu} \pi^{\rho\sigma} + \frac{1}{2\sqrt{h}}\Pi_{\mu}\Pi^{\mu} + 
\frac{\sqrt{h}}{4} F_{\mu\nu} F^{\mu\nu}  - \sqrt{h}R    = 0 \mbox{ } ,
\label{EMHam}
\eeq
i.e. the Einstein--Maxwell Hamiltonian constraint, ${\cal H}^{\sA}$ .  
Variation with respect to $\Psi$ gives
\beq
G(x^{\mu}) = \frac{\pa\bar{\fL}_{\sA}}{\pa\dot{\Psi}} = D_{\mu}\Pi^{\mu} = \pa_{\mu}\Pi^{\mu} \mbox{ } , 
\eeq
and the FESH conditions 
$\pa\bar{\fL}_{\sA}/\pa\dot{\Psi}|_{\lambda_i} = 0 = 
\pa\bar{\fL}_{\sA}/\pa\dot{\Psi}|_{\lambda_f}$, 
so $G(x^{\mu}) = \pa\bar{\fL}_{\sA}/\pa\dot{\Psi} = 0$ at the end spatial hypersurface, but 
$G$ is hypersurface-independent, so $G = 0$ everywhere.  
Thus one arrives at 
\beq
D_{\mu} {\Pi^{\mu} }  = 0 \mbox{ } , 
\label{Gau}
\eeq
which is the electromagnetic Gauss constraint, ${\cal G}$, of Einstein--Maxwell theory.   
Variation with respect to $F_{\mu}$ gives (modulo a Gauss constraint term) also as a secondary constraint 
\beq
E_{\gamma}(x^{\mu}) = \frac{\pa\bar{\fL}_{\sA}}{\pa\dot{F^{\gamma}}} = 
-2 D_{\nu} {\pi_{\mu} }^{\nu} -\Pi^{\delta}\{D_{\delta}A_{\gamma} - D_{\gamma}A_{\delta}\} 
\eeq
and the FESH conditions 
$\pa\bar{\fL}_{\sA}/\pa\dot{F}^{\gamma}|_{\lambda_i} = 0 = 
\pa\bar{\fL}_{\sA}/\pa\dot{F}^{\gamma}|_{\lambda_f}$, 
so $E_{\gamma}(x^{\mu}) = \pa\bar{\fL}_{\sA}/\pa\dot{F}^{\gamma} = 0$ at the end spatial hypersurface, 
but $E_{\gamma}$ is hypersurface-independent, so $E_{\gamma} = 0$ everywhere.
Thus one arrives at  the momentum constraint of 
Einstein--Maxwell theory, ${\cal H}_{\mu}$,  
\beq
-2 D_{\delta} {\pi_{\gamma} }^{\delta}  = \Pi^{\delta}\{D_{\delta}A_{\gamma} - D_{\gamma}A_{\delta}\} 
\mbox{ } .   
\label{EMMom}
\eeq
This formulation's equations of motion $\dot{\pi}^{\mu\nu} = \pa \bar{\fL}_{\sA}/\pa h_{\mu\nu}$, 
$\dot{\Pi}^{\mu} = \pa\bar{\fL}_{\sA}/\pa A_{\mu}$ propagate the above constraints without giving rise 
to any further ones.

Taking the Lagrangian form of (\ref{Ham}), one can now solve for $\dot{I}$: 
\beq
\dot{I} = \pm \frac{1}{2}\sqrt{ \frac{\fT + \fT_A}{R + \fU_A}} 
\mbox{ } 
\eeq
(and take the + sign by convention of direction of forward march of time), from which $\dot{I}$ is 
straightforwardly algebraically eliminable from the action (\ref{Lemcssplit}) 
by Routhian reduction.  
Thus one recovers the `BFO--A' action 
\beq
\fI_{\sB\sF\sO{-}\sA}[h_{\mu\nu}, \dot{h}_{\mu\nu}, A_{\mu}, \dot{A}_{\mu}, \dot{F}_{\mu}, \dot{\Psi}] = 
\int\d\lambda\int\d^3x\bar{\fL}_{\sB\sF\sO{-}\sA}
(\dot{h}_{\mu\nu}, \dot{A}_{\mu}; h_{\mu\nu}, A_{\mu}, \dot{F}_{\mu}, \dot{\Psi}] \equiv 
\int\d\lambda\d^3x\sqrt{\{\fT + \fT_{\sA}\}\{R + \fU_{\sA}\}} =
\eeq
\beq
\int\d\lambda\int\d^3x\sqrt{h}
\sqrt{             \left\{           \stackrel{    \mbox{$\underline{G}^{\mu\nu\rho\sigma} 
\{\dot{h}_{\mu\nu}     - \pounds_{\dot{F}}h_{\mu\nu}    \}
\{\dot{h}_{\rho\sigma} - \pounds_{\dot{F}}h^{\rho\sigma}\} \mbox{ } + $    }} 
                           {    2 h^{\mu\nu}
\{\dot{A}_{\mu} - \pounds_{\dot{F}}A_{\mu} - \pa_{\mu} \dot{\Psi}\}
\{\dot{A}_{\nu} - \pounds_{\dot{F}}A_{\nu} - \pa_{\nu} \dot{\Psi}\}    }         \right\}
\left\{ 
R - \frac{1}{4}F_{\mu\nu}F^{\mu\nu} 
\right\}             } \mbox{ } .
\label{BFOA}
\eeq
This is my formulation's counterpart and equivalent of BSW's multiplier elimination procedure \cite{BSW},  
thus revealing \cite{BG70}'s {\it ``explicit use is made of multiplier status"} to be an alternative 
rather than an obligation.

Taking this action as starting point, the momenta are
\beq
\pi^{\mu\nu} = \frac{\pa\bar{\fL}_{\sB\sF\sO{-}\sA}}{\pa \dot{h}_{\mu\nu}} = 
\sqrt{\frac{R + \fU_{\sA}}{\fT + \fT_{\sA}}}G^{\mu\nu\rho\sigma} 
\{\dot{h}_{\rho\sigma} - \pounds_{\dot{F}}h_{\mu\nu}\}
\mbox{ } , \mbox{ } 
\Pi^{\mu} = \frac{\pa\bar{\fL}_{\sB\sF\sO{-}\sA}}{\pa \dot{A}_{\mu}} = 
\sqrt{h}\sqrt{\frac{R + \fU_{\sA}}{\fT + \fT_{\sA}}}h^{\mu\nu}
\{\dot{A}_{\nu} - \pounds_{\dot{F}}A_{\nu} - \pa_{\nu}\dot{\Psi}\} 
\eeq
-- now one does not have not a priori an instant $I$ of which the rate appears in the equations, but 
rather an emergent quantity $\frac{1}{2}\sqrt{\frac{R + \sfU_{\tA}}{\sfT + \sfT_{\tA}}}$.  
Then there is a primary constraint 
\beq
\stackrel{\mbox{$G_{\mu\nu\rho\sigma}\pi^{\mu\nu}\pi^{\rho\sigma}$}}
         { + 2 h^{\mu\nu}\Pi^{\mu}\Pi^{\nu} }  
=
\stackrel{\mbox{$G_{\mu\nu\rho\sigma}G^{\mu\nu\theta\phi}
\sqrt{\frac{R+\sfU_{\tA}}{\sfT+\sfT_{\tA}}}
\{\dot{h}_{\theta\phi}-\pounds_{\dot{F}}h_{\theta\phi}\}
G^{\rho\sigma\eta\zeta }
\sqrt{           \frac{   R + \sfU_{\tA}  }{  \sfT + \sfT_{\tA}     }          }
\{\dot{h}_{\eta\zeta}-\pounds_{\dot{F}}h_{\eta\zeta}\}+$}}
{\mbox{$h_{\mu\nu}h^{\mu\rho} 
\sqrt{           \frac{   R + \sfU_{\tA}  }{  \sfT + \sfT_{\tA}     }          }
\{\dot{A}_{\rho}-\pounds_{\dot{F}}A_{\rho}-\pa_{\rho}\dot{\Psi}\}
h^{\nu\sigma}
\sqrt{           \frac{   R + \sfU_{\tA}  }{  \sfT + \sfT_{\tA}     }          }
\{\dot{A}_{\sigma}-\pounds_{\dot{F}}A_{\sigma}-\pa_{\sigma}\dot{\Psi}\}           $}} 
= \sqrt{h}\frac{R + \fU_{\sA}}{\fT + \fT_{\sA}}\{\fT + \fT_{\sA}\} = \sqrt{h}\{R + \fU_{\sA}\}
\mbox{ } 
\eeq
(by the definition of momentum, and using that $G_{\alpha\beta\gamma\delta}$ is the inverse of 
$G^{\alpha\beta\gamma\delta}$), the Hamiltonian constraint.  
Variation with respect to $\Psi$ gives 
\beq
G(x^{\mu}) = \frac{\pa\bar{\fL}_{\sB\sF\sO{-}\sA}}{\pa\dot{\Psi}} = 
D_{\mu}\Pi^{\mu} = \pa_{\mu}\Pi^{\mu} \mbox{ } , 
\eeq
and the FESH conditions 
$\pa\bar{\fL}_{\sB\sF\sO{-}\sA}/\pa\dot{\Psi}|_{\lambda_i} = 0 = 
\pa\bar{\fL}_{\sB\sF\sO{-}\sA}/\pa\dot{\Psi}|_{\lambda_f}$, 
so $G(x^{\mu}) = \pa\bar{\fL}_{\sB\sF\sO{-}\sA}/\pa\dot{\Psi} = 0$ at the end spatial hypersurface, 
but $G$ is hypersurface-independent, so $G = 0$ everywhere.
Variation with respect to $F_{\mu}$ gives (modulo a Gauss constraint term) the secondary constraint 
\beq
E_{\gamma}(x^{\mu}) = \frac{\pa\bar{\fL}_{\sB\sF\sO{-}\sA}}{\pa \dot{F}^{\gamma}} = 
-2 D_{\nu}{\pi^{\mu}}_{\nu} - \Pi^{\delta}\{D_{\delta}A_{\gamma} - D_{\gamma}A_{\delta}\} 
\eeq
and the FESH conditions 
$\pa\bar{\fL}_{\sB\sF\sO{-}\sA}/\pa\dot{F^{\gamma}}|_{\lambda_i} = 0 = 
\pa\bar{\fL}_{\sB\sF\sO{-}\sA}/\pa\dot{F^{\gamma}}|_{\lambda_f}$, so $E_{\gamma}(x^{\mu}) = 
\pa\bar{\fL}_{\sB\sF\sO{-}\sA}/\pa\dot{F}_{\gamma} = 0$ at the end spatial hypersurface, 
but $F_{\gamma}$ is hypersurface-independent, so $F_{\gamma} = 0$ everywhere.
Thus one arrives at 
\beq
-2 D_{\nu} {\pi_{\mu} }^{\nu}  = \Pi^{\delta}\{D_{\delta}A_{\gamma} - D_{\gamma}A_{\delta}\}  \mbox{ } , 
\label{Mom}
\eeq
the momentum constraint of GR, ${\cal H}_{\mu}$.
The BFO--A equations of motion $\dot{\pi}^{\mu\nu} = \pa \bar{\fL}_{\sB\sF\sO{-}\sA}
/\pa h_{\mu\nu}$ propagate these constraints without giving rise to any further ones.

Thus in this scheme temporal relationalism as implemented by manifest reparametrization invariance 
without extraneous variables gives the Hamiltonian constraint of GR and spatial relationalism as regards 
3-diffeomorphisms as implemented by arbitrary G-frame cyclic coordinate velocity corrections to the 
metric velocities gives the GR momentum constraint.  
In fact, even if a number of features of $\fL_{\sB\sF\sO{-}\sA}$ are not known, if it is built according 
to temporal relationalism, the Dirac procedure is applied and consistency is demanded, 
$\fI_{\sB\sF\sO{-}\sA}$ emerges as one of very few possibilities \cite{RWR, AB, Lan, Phan}. 
The above very clear way of obtaining the GR constraints (and lack of prior knowledge of the form of the 
GR equations to do so) and the accommodability into this scheme of a sufficiently wide range of fundamental 
matter fields to describe nature at the classical level \cite{AB, Lan, Phan} make (\ref{BFOA}) itself 
interesting as a starting point for gravitational physics.
In fact,  starting with temporal relationalism (and a configuration space of spatial 3-metrics) alone  
suffices \cite{OM02San}, as that produces ${\cal H}$ as above, which then happens to dictate ${\cal H}$ 
and ${\cal G}$ as integrability conditions (see also \cite{MT72}), whose subsequent appending or encoding leads to the recovery of 
(\ref{BFOA}).\foo{Strictly 
speaking, these references consider the BSW formulation counterpart of 
these results; the BFO-A counterpart to these results is a new result of the present paper.}  
%
Though this may be considered to involve luck, as, in other cases \cite{OM02San}, relying on 
integrability can cause constraints to get `missed out'.

\section{Conclusion}

In this paper, I have explained that variation with respect to auxiliary variables is free end point 
(FEP) in finite cases or free end spatial hypersurface (FESH) in field-theoretic cases. 
This makes a difference to the outcome of varying cyclic coordinates if they are auxiliary, whereby 
interpreting an auxiliary present in one's action as a cyclic velocity is equivalent to interpreting it 
as a multiplier coordinate  
Thus one has an alternative to the ADM split: instead of lapse $\alpha$ and shift $\beta_{\mu}$ 
multiplier coordinates, one has instant $I$ and grid $F_{\mu}$ cyclic coordinates, which are related 
to the $\alpha$ and $\beta_{\mu}$ by $\dot{I} = \alpha$ and $\dot{F}_{\mu} = \beta_{\mu}$.    
Also one has an alternative to the usual electric-magnetic split formulation: instead of 
an electric potential multiplier coordinate $\Phi$, one has a flux cyclic coordinate $\Psi$ that is 
related to it by $\dot{\Psi} = \Phi$.

Further principles of dynamics manipulations continue to give the same answers in the new scheme.  
For example, Routhian reduction now works out to be equivalent to what was previously multiplier 
elimination.  
E.g. the new scheme has a Routhian reduction instant-velocity elimination counterpart of 
Baierlein, Sharp and Wheeler's multiplier elimination of the lapse.  
As another example, this scheme supplants Hamiltonians $\fH$ by `almost Hamiltonians' $\fA$ which 
depend on positions and momenta for the usual variables but on postions and velocities for manifestly 
pure-frame variables. 
There is then no difference in constraints arising from the multiplier picture's $\fH$ and the 
cyclic coordinate picture's $\fA$ in the position--momentum language for the theories in this paper 
(for which the auxiliary variables, however represented, do not enter the position-momentum form 
of the constraints).    
Nor is there any difference in the thin sandwich conjecture (solving the Lagrangian form of 
${\cal H}_{\mu}$ with matter struck out for $\beta_{\mu}$ \cite{TSC} or now for $\dot{F}_{\mu}$, 
or the Lagrangian form of ${\cal H}_{\mu}$ and ${\cal G}$ for $\beta_{\mu}$ and $\Phi$ \cite{Giulini} 
or now for $\dot{F}_{\mu}$ and $\dot{\Psi}$).

One advantage of this paper's formulation is that it is more manifestly temporally relational than ADM's. 
It is true that the instant-less BFO-A formalism is `even more' manifestly temporally relational in 
Barbour's sense of not containing an extraneous time variable, but, on the ther hand, thinking along 
the lines of this paper's formalism and then eliminating the extraneous time variable permits the 
relational program to go beyond its current spatially compact without boundary setting to cases 
with boundaries or that are open.

The cyclic velocity picture gives a different account of how the frozen formalism problem of GR 
(see e.g. \cite{I93, Kuchar, B94II, EOT}) arises.    
\beq
{\cal H} = \frac{\pa\fO}{\pa \dot{I}} = - \frac{\pa\fL}{\pa \dot{I}} \equiv - \Pi^{\fI} \mbox{ }   
\eeq
by, respectively, trivial direct computation, the chain rule applied to the Legendre transformation 
relation and the definition of momentum. 
Thus 
\beq
\Pi^{\sI} + {\cal H} = 0 \mbox{ } .  
\eeq
Similarly, 
\beq
\Pi^{\sF}_{\mu} + {\cal H}_{\mu} = 0 \mbox{ } .  
\eeq
At this stage, these contain linear momenta, so, in particular, the first quadratic equation has no 
trace of the frozen formalism.  
It would amount to 
\beq
i\hbar\frac{\pa|\Psi>}{\pa I} = \widehat{{\cal H}}|\Psi> 
\eeq
at the quantum level, which is an (instant, or proper,) time-dependent Schr\"{o}dinger 
equation rather than a stationary equation like the Wheeler--DeWitt equation.  
However, one has also the FESH condition $\Pi^{\sI} = 0$, which precisely wipes out the 
linear momentum in the crucial, quadratic constraint, leaving 
\beq
{\cal H} = 0 \mbox{ } .  
\eeq
At the quantum level this amounts to the usual stationary Wheeler-DeWitt equation 
\beq
\widehat{{\cal H}}\Psi = 0  \mbox{ } .
\eeq
However, in cases/simplified models in which there is a privileged background, the situation here is, 
very transparently, that FEP is then inappropriate, whereby some kind of time-dependent Schr\"{o}dinger 
equation survives.  
As many quantum schemes proceed via the position--momentum form of the constraints that is unchanged 
by passing to this paper's formalism, these schemes are likewise unchanged.  
I leave investigation of whether this article's change has any effect on any other quantization 
procedures for a future occasion.

Finally, I explained how auxiliary variables can be more general than either multiplier coordinates or 
the velocities corresponding to cyclic coordinates, which affords further extension of how gauge 
theories are treated in this paper.  
This taking one further afield than standard formulations of standard theories of physics (e.g. these 
more general cases apply to \cite{ABFO, ABFKO}), I present these in a second paper \cite{ADMII}.

\mbox{ } 

\noindent{\bf\large Acknowledgements}

\mbox{ }

\noindent I thank Dr Julian Barbour, Prof Niall \'{O} Murchadha, Dr Brendan Foster and Dr Bryan Kelleher 
for past collaboration, and them and Dr Jeremy Butterfield, Prof William Unruh and Mr Eyo Ita for discussions, 
and Peterhouse for funding.  

\mbox{ }

\noindent{\bf{\large Appendix: ADM and BSW-type formalisms of Einstein--Maxwell theory}}  

\mbox{ }

\noindent The conventional 3 + 1 split action for Einstein--Maxwell theory is
$$
\fI_{\sfA\sfD\sfM}[h_{\mu\nu}, \dot{h}_{\mu\nu}, A_\mu, \dot{A}_{\mu}, \beta_{\mu}, \alpha, \Phi] = 
\int\d\lambda\int\d^3x\bar{\fL}_{\sfA}
(\dot{h}_{\mu\nu}, \dot{A}_{\mu}, \alpha; h_{\mu\nu}, A_\mu, \beta_{\mu}, \Phi]  = 
\int\d\lambda\int\d^3x\sqrt{h}\alpha
\left\{
\frac{\fT + \fT_{\sA}}{\alpha^2} + R - \frac{1}{4}F_{\mu\nu}F^{\mu\nu}
\right\}
$$
$$
=  \int\d\lambda\int\d^3x\sqrt{h}\alpha 
\left\{
\frac{\underline{G}^{\mu\nu\rho\sigma}
\{\dot{h}_{\mu\nu} - \pounds_{\beta}h_{\mu\nu}\}
\{\dot{h}_{\rho\sigma} - \pounds_{\beta}h_{\rho\sigma}\}}{4{\alpha}^2} + 
\frac{        h^{\mu\nu}\{\dot{A}_{\mu} - \pounds_{\beta} A_{\mu} - \pa_{\mu}{\Phi}\}
              \{\dot{A}_{\nu }- \pounds_{\beta} A_{\nu} - \pa_{\nu}{\Phi}\}        }
               {       2{\alpha}^2        }\right. 
$$
\beq
\left.
- \frac{1}{4}F_{\mu\nu}F^{\mu\nu} + R 
\right\} \mbox{ } .  
\label{Lemcssplit}
\eeq
Here, $\alpha$ is the lapse, $\beta^{\mu}$ is the shift and $\Phi$ is the electric potential; 
these are all taken here to be Lagrange multipliers.  
See footnote 7 for the rest of the notation.

From this, the conjugate momenta are 
\beq
\pi^{\mu\nu} = \frac{  \pa\bar{{\fL}}_{\sA\sD\sM}  }{\pa\dot{h}_{\mu\nu}} = 
\frac{ 1 }{ 2\alpha }G^{\mu\nu\rho\sigma}
\{\dot{h}_{\mu\nu} - \pounds_{\beta}h_{\mu\nu}\}
\mbox{ } \mbox{ and } \mbox{ }
\Pi^{\mu} \equiv 
\frac{\pa {\fL}}{\pa \dot{A}_{\mu}} = 
\frac{\sqrt{h}}{\alpha}h^{\mu\nu}\{\dot{A}_{\mu} - \pounds_{\beta}A_{\mu} - \pa_{\mu}{\Phi}\} 
\mbox{ } . 
\eeq
Variation with respect to $\alpha$ gives the Einstein--Maxwell Hamiltonian constraint, ${\cal H}^{\sA}$: 
\beq
G_{\mu\nu\rho\sigma}\pi^{\mu\nu} \pi^{\rho\sigma} + \frac{1}{2\sqrt{h}}\Pi_{\mu}\Pi^{\mu} + 
\frac{\sqrt{h}}{4} F_{\mu\nu} F^{\mu\nu}  - \sqrt{h}R    = 0 \mbox{ } .
\label{EMHam2}
\eeq 
Variation with respect to $\Phi$ gives the Gauss constraint, ${\cal G}$: 
\beq
D_{\mu} {\Pi^{\mu} }  = 0 \mbox{ } . 
\label{Gau2}
\eeq
Variation with respect to $F_{\mu}$ gives (modulo a Gauss constraint term) 
the Einstein--Maxwell momentum constraint ${\cal H}^{\sA}_{\mu}$:  
\beq
-2 D_{\delta} {\pi_{\gamma} }^{\delta}  = \Pi^{\delta}\{D_{\delta}A_{\gamma} - D_{\gamma}A_{\delta}\}  
\mbox{ } . 
\label{EMMom2}
\eeq
One then finds that propagation of these by the ADM equations of motion for Einstein--Maxwell theory,  
$\dot{\pi}^{\mu\nu} = \pa \bar{\fL}_{\sB\sS\sW}
/\pa h_{\mu\nu}$, $\dot{\Pi}^{\mu} = \pa \bar{\fL}_{\sB\sS\sW}
/\pa A_{\mu}$, gives no further constraints.

Using the Lagrangian variables version of the $\alpha$-multiplier equation,  
\beq
R - \frac{\fT}{4\alpha^2} = 0 
\mbox{ } \Rightarrow \mbox{ }  
\alpha = \pm \frac{1}{2}\sqrt{ \frac{\fT}{R}} 
\mbox{ } 
\eeq
(and taking the + sign by convention of direction of forward march of time), $\alpha$ is 
straightforwardly algebraically eliminable from the ADM action 
giving the Baierlein--Sharp--Wheeler (BSW)-type action for Einstein--Maxwell theory: 
\beq
\fI_{\sB\sS\sW}[h_{\mu\nu}, \dot{h}_{\mu\nu}, A_\mu, \dot{A}_{\mu}, \beta_{\mu}, \Phi] = 
\int\d\lambda\int\d^3x \bar{\fL}_{\sB\sS\sW}(\dot{h}_{\mu\nu}, \dot{A}_{\mu}; h_{\mu\nu}, A_\mu, 
\beta_{\mu}, \Phi] = \int\d\lambda\d^3x\sqrt{\{\fT + \fT_{\sA}\}\{R + \fU_{\sA}\}} =
\eeq
\beq
\int\d\lambda\int\d^3x\sqrt{h}
\sqrt{
\left\{
\stackrel{\mbox{$\underline{G}^{\mu\nu\rho\sigma} + 
\{\dot{h}_{\mu\nu} - \pounds_{\beta}h_{\mu\nu}\}
\{\dot{h}_{\rho\sigma} - \pounds_{\beta}h^{\rho\sigma}\} +$}} 
{2 h^{\mu\nu}\{\dot{A}_{\mu} - \pounds_{\beta}A_{\mu} - \pa_{\mu}{\Phi}\}
            \{\dot{A}_{\nu} - \pounds_{\beta}A_{\nu} - \pa_{\nu}{\Phi}\}}         
\right\}
\left\{
R - \frac{1}{4}F_{\mu\nu}F^{\mu\nu}
\right\}                          }  
\mbox{ } .
\label{4}
\eeq

From this starting point, the conjugate momenta are then 
\beq
\pi^{\mu\nu} = \frac{\pa\bar{\fL}_{\sB\sS\sW}}{\pa \dot{h}_{\mu\nu}} = \sqrt{\frac{R + \fU_{\sA}}
{\fT + \fT_{\sA}}}
G^{\mu\nu\rho\sigma}\{\dot{h}_{\rho\sigma} - \pounds_{\beta}h_{\rho\sigma} \}
\mbox{ } \mbox{ and } \mbox{ } 
\Pi^{\mu} \equiv \frac{\pa\bar{\fL}_{\sB\sS\sW}}{\pa \dot{A}_{\mu}} = 
\sqrt{h}\sqrt{\frac{R + \fU_{\sA}}{\fT + \fT_{\sA}}}
h^{\mu\nu}\{\dot{A}_{\nu} - \pounds_{\beta}A_{\nu} - \pounds_{\nu}\Phi\} \mbox{ } .  
\eeq
Then there arises as a primary constraint the Hamiltonian 
constraint of Einstein--Maxwell theory (\ref{EMHam2}). 
Variation with respect to $\Phi$ gives the Gauss constraint (\ref{Gau2}).
Variation with respect to $F_{\mu}$ gives the Einstein--Maxwell momentum constraint (\ref{EMMom2})  
(again modulo a Gauss constraint term).  
One then finds that propagation of these by the BSW equations of motion 
$\dot{\pi}^{\mu\nu} = \pa \bar{\fL}_{\sB\sS\sW}
/\pa h_{\mu\nu}$,
$\dot{\Pi}^{\mu} = \pa \bar{\fL}_{\sB\sS\sW}
/\pa A_{\mu}$, again gives no further constraints.



\begin{thebibliography}{99}

\footnotesize

\bibitem{ADM}                 R. Arnowitt, S. Deser and C.W. Misner, in \it{Gravitation: an
                              Introduction to Current Research} \normalfont ed L. Witten (Wiley, New York 1962).

\bibitem{MTW}                 C.W. Misner, K. Thorne and J.A Wheeler, {\it Gravitation} (Freedman, San Francisco 1973).


\bibitem{WheelerGRT}          J.A. Wheeler, in {\it Relativity, Groups and Topology} ed. C. DeWitt \& B. DeWitt 
                              (Gordon and Breach, New York and London 1964).

\bibitem{DeWitt}              B.S. DeWitt, Phys. Rev. {\bf 160} 1113 (1967). 

\bibitem{Lanczos}             C. Lanczos, {\it The Variational Principles of Mechanics} (University of Toronto Press, Toronto 1949).

\bibitem{Battelle}            J.A. Wheeler, in  {\it Battelle Rencontres: 1967 Lectures in Mathematics 
                              and Physics} ed. C. DeWitt and J.A. Wheeler (Benjamin, New York 1968). 

\bibitem{I93}                 C.J. Isham, in {\it Integrable Systems, Quantum Groups and Quantum Field 
                              Theories} ed. L.A. Ibort and M.A. Rodr\'{i}guez (Kluwer, Dordrecht 1993), 
                              gr-qc/9210011.

\bibitem{Kuchar}              K.V. Kucha\v{r}, {\it Quantum Gravity 2: A Second Oxford Symposium} 
                              ed. in C.J. Isham, R. Penrose and D. Sciama (Clarendon Press, Oxford 1981); 
                              {\it Conceptual Problems of Quantum Gravity} 
                              ed. A. Ashtekar and J. Stachel (Birkh{\"a}user, Boston 1991);      
                              in {\it Proceedings of the 4th Canadian Conference on 
                              General Relativity and Relativistic Astrophysics} ed. G. Kunstatter, 
                              D. Vincent and J. Williams (World Scientific, Singapore 1992).   

\bibitem{BSG}                 T.W. Baumgarte and S.L. Shapiro,  Phys. Rept. {\bf 376} 41 (2003), gr-qc/0211028; 
                              E. Gorghoulon, arXiv:gr-qc/0703035.  

\bibitem{Thorne}              See e.g. K.S. Thorne et al, in {\it Black Holes: The Membrane Paradigm} ed. K.S Thorne, R.H. Price and 
                              D.A. MacDonald (Yale University Press, New Haven and London 1986) p 68 and p 70; 
                              there is also brief mention in a specific example in \cite{WheelerGRT}.  


\bibitem{HKT}                 S.A. Hojman, K.V. Kucha\v{r} and C. Teitelboim, Ann. Phys. N.Y. {\bf 96} 
                              88 (1976).

\bibitem{Leibniz}             See e.g. {\it The Leibniz--Clark Correspondence}, ed. H.G. Alexander (Manchester 1956). 

\bibitem{Mach}                E. Mach, {\it Die Mechanik in ihrer Entwickelung, Historisch-kritisch dargestellt} (J.A. Barth, Leipzig 1883).  
                              The Enlish translation is \it The Science of Mechanics: A Critical and Historical Account of its Development \normalfont (Open Court, La Salle, Ill. 1960).    

\bibitem{BB82}                J.B. Barbour and B. Bertotti, Proc. Roy. Soc. Lond. {\bf A382} 295 (1982).
                              J.B. Barbour, in {\it Quantum Concepts in Space and Time} 
                              ed. R. Penrose and C.J. Isham (Oxford University Press, Oxford 1986).      

\bibitem{B94I}                J.B. Barbour, Class. Quantum Grav. \bf 11 \normalfont 2853 (1994).

\bibitem{RWR}                 J.B. Barbour, B.Z. Foster and N. \'{O} Murchadha, Class. Quantum Grav. \bf 19 \normalfont 3217 (2002), gr-qc/0012089.

\bibitem{Lan}                 E. Anderson, in {\it General Relativity Research Trends, Horizons in World 
                              Physics} {\bf 249} ed. A. Reimer (Nova, New York 2005), gr-qc/0405022. 

\bibitem{B03}                 J.B. Barbour, Class. Quantum Grav. \textbf{20}, 1543 (2003), gr-qc/0211021.    

\bibitem{BSW}                 R.F. Baierlein, D. Sharp and J.A. Wheeler, Phys. Rev. {\bf 126} 1864 (1962).

\bibitem{ABFO}                E. Anderson,  J.B. Barbour, B.Z. Foster and N. \'{O} Murchadha, Class. Quantum Grav. {\bf 20} 157 (2003), gr-qc/0211022.

\bibitem{Phan}                E. Anderson, Studies in History and Philosophy of Modern Physics, {\bf 38} 15 (2007), gr-qc/0511070.

\bibitem{CHFoxBM}             R. Courant and D. Hilbert, {\it Methods of Mathematical Physics} Vol. 2 (John Wiley and Sons, Chichester 1989); 
                              C. Fox, {\it An Introduction to the Calculus of Variations} (Oxford University Press, London 1950); 
                              U. Brechtken-Manderscheid {\it Introduction to the Calculus of Variations} 
                              (T.J. Press, Padstow, Cornwall 1991: English translation of 1983 German text).  

\bibitem{Dirac}               P.A.M. Dirac, \it Lectures on Quantum Mechanics \normalfont (Yeshiva University, New York 1964).

\bibitem{EOT}                 J.B. Barbour, {\it The End of Time} (Oxford University Press, Oxford 1999).

\bibitem{ERPM}                J.B. Barbour and L. Smolin, unpublished, dating from 1989; 
                              L. Smolin, in {\it Conceptual Problems of Quantum Gravity}, ed. A. Ashtekar and J. Stachel (Birkh\"{a}user, Boston  1991); 
                              C. Rovelli, p. 292 in {\it Conceptual Problems of Quantum 
                              Gravity}, ed. A. Ashtekar and J. Stachel (Birkh\"{a}user, Boston  1991); 
                              D. Lynden-Bell, in \it Mach's principle: From Newton's Bucket to Quantum Gravity\normalfont, 
                              ed. J.B. Barbour and H. Pfister (Birkh\"{a}user, Boston 1995);  
                              L.\'{A} Gergely, Class. Quantum Grav. {\bf 17} 1949 (2000), gr-qc/0003064; 
                              L.\'{A} Gergely and M. McKain, Class. Quantum Grav. {\bf 17} 1963 (2000), gr-qc/0003065; 
                              C. Kiefer, {\it Quantum Gravity} (Clarendon, Oxford 2004); 
                              E. Anderson, Class. Quant. Grav. {\bf 23} 2469 (2006), gr-qc/0511068;
                              {\bf 24} 2935 (2007), gr-qc/0611007;
                              {\bf 24} 2971 (2007), gr-qc/0611008.
  
\bibitem{ERPMSRPM}            E. Anderson, AIP Conf. Proc. {\bf 861} 285 (2006), gr-qc/0509054; 
                              Class. Quant. Grav. {\bf 23} 2491 (2006), gr-qc/0511069; 
                              Class. Quantum Grav. {\bf 24} 5317 (2007), gr-qc/0702083;  
                              arXiv:0706.3934;
                              arXiv:0709.1892;
                              ``Quantum Mechanics in Triangleland", forthcoming; 
                              ``Dynamics without Dynamics: Angular Momentum Exchange in Classical and Quantum Triangleland", forthcoming;
                              2 papers on Semiclassical Quantum Gravity, forthcoming;  
                              ``What is the Distance between two Shapes?", forthcoming.    

\bibitem{SRPM}                J.B. Barbour, in {\it Decoherence and Entropy in Complex Systems 
                              (Proceedings of the Conference DICE, Piombino 2002} ed. H. -T. Elze 
                              (Springer Lecture Notes in Physics 2003), gr-qc/0309089.     

\bibitem{HTbook}              M. Henneaux and C. Teitelboim, {\it Quantization of Gauge Systems} (Princeton University Press, Princeton, New Jersey 1992).

\bibitem{MT72}                V. Moncrief and C. Teitelboim, Phys. Rev. {\bf D6} 966 (1972).

\bibitem{OM02San}             N. \'{O} Murchadha, Int. J. Mod. Phys. {\bf A 20} 2717 (2002); 
                              E. Anderson, Gen. Rel. Grav. {\bf 36} 255, gr-qc/0205118.

\bibitem{ABFKO}               E. Anderson, J.B. Barbour, B.Z. Foster, B. Kelleher and N. \'{O} Murchadha, Class. Quantum Grav {\bf 22} 1795 (2005), gr-qc/0407104.  

\bibitem{CG}                  J.B. Barbour and N. \'{O} Murchadha, , gr-qc/9911071.

\bibitem{York}                J.W. York, Phys. Rev. Lett. {\bf 28} 1082 (1972); J. Math. Phys. {\bf 14} 456 (1973).

\bibitem{ADMII}               E. Anderson, ``New Interpretation of Variational Principles for Gauge Theories.  
                              II.  General Auxiliary Coordinate in Conformogeometrodynamics.", forthcoming.   
       
\bibitem{BG70}                D.R. Brill and R.H. Gowdy, Rep. Prog. Phys. {\bf 33} 413 (1970).  

\bibitem{AB}                  J.B. Barbour, B.Z. Foster and N. \'{O} Murchadha, v1 of gr-qc/0012089;
                              E. Anderson and J.B. Barbour, Class. Quantum Grav. \bf 19 \normalfont 3249 (2002), gr-qc/0201092;
                              E. Anderson, Phys. Rev. {\bf D68} 104001 (2003), gr-qc/0302035;
                              N. \'{O} Murchadha, gr-qc/0305038;
                              ``Geometrodynamics: spacetime or space?" (Ph.D. Thesis, University of London 2004), gr-qc/0409123;
                              ``Does Relationalism alone control Geometrodynamics with Sources?", 
                              arXiv:0711.0285.   


\bibitem{B94II}               J.B. Barbour, Class. Quantum Grav. \bf 11 \normalfont 2875 (1994).



\bibitem{TSC}                 E.P. Belasco and H.C. Ohanian, J. Math. Phys. {\bf 10}, 1053 (1969);

                              R. Bartnik and G. Fodor, Phys. Rev. {\bf D48}, 3596 (1993). 

\bibitem{Giulini}             D. Giulini, J. Math. Phys. {\bf 40}, 1470 (1999).

\bibitem{BarbModPhys}         E. Anderson, ``Barbour's Relationalism and Modern Theoretical Physics", forthcoming.

   






\end{thebibliography}
\end{document}